\RequirePackage{ifpdf}

\documentclass[prd,showpacs,aps,reprint,nofootinbib]{revtex4-1}

\usepackage[utf8]{inputenc}
\usepackage[T1]{fontenc}
\usepackage{ae,aecompl}
\usepackage{bm}
\usepackage{latexsym}
\usepackage{dcolumn}
\usepackage{amsfonts,amssymb}
\usepackage{graphicx,epsfig}
\usepackage{sidecap}
\usepackage[style=base]{caption, subcaption}
\usepackage[labelfont={bf},textfont={normalfont},justification=raggedright]{caption}
\usepackage{psfrag}
\usepackage{amsmath,amssymb,multirow,array,mathtools,rotating}
\usepackage{empheq}
\usepackage{tabularx}
\usepackage{float}

\usepackage{tcolorbox}
\usepackage{xcolor}
\usepackage{changepage}

\usepackage{natbib}
\usepackage{silence}
\usepackage{hyperref} 
\usepackage{url}
\hypersetup{
	colorlinks=true,	
	linkcolor=blue,		
	citecolor=red,		
	filecolor=blue,		
	urlcolor=blue		
}

\usepackage{cleveref}
\usepackage{soul, todonotes}

\newcommand{\bbar}{\mathchar'26\mkern-9mu b}

\usepackage{tocbasic} 
\usepackage[titletoc,title]{appendix}
\numberwithin{equation}{section}
\begin{document}
	\pagestyle{plain}
	\title{Interaction Between AdS Black Hole Molecules}
	\author{Suvankar Dutta}
	\email[]{suvankar@iiserb.ac.in}
	\affiliation{Indian Institute of Science Education \&\ Research Bhopal, Bhopal Bypass, Bhopal 462066, India}
	\author{Gurmeet Singh Punia} 
	\email[]{gurmeet17@iiserb.ac.in}
	\affiliation{Indian Institute of Science Education \&\ Research Bhopal, Bhopal Bypass, Bhopal 462066, India}
	\begin{abstract}
		We take a bottom-up approach to find the interaction potential between the $AdS$ black hole molecules under mean-field approximation. We start with the equation of state of dyonic $AdS$ black holes in fixed charge ensemble and use the method of classical cluster expansion to find the mean-field potential. We show that the Lennard-Jones (LJ) potential is a feasible choice to describe the equation of state. The LJ potential describes a two-body interaction. There exists a critical distance $r_0$ such that two interacting particles repel (attract) each other for $r < r_0$ ($r > r_0$). We compute the value of $r_0$ for dyonic $AdS$ black holes and compare the result obtained from the Ruppeiner scalar curvature. Our analysis shows how the electric (and magnetic) charge affects the interaction between black hole molecules.	
	\end{abstract}
\maketitle 
\tableofcontents

\section{Introduction and summary}\label{Sec:intro}
Black holes are believed to be thermal objects \cite{Hawking:1976_BH_thermo}. Different macroscopic variables of black holes are identified with those of standard thermodynamics and the relations between them resemble the laws of thermodynamics \cite{Hawking:19973_BHT_Four_laws}. For example, the area of the event horizon, the ADM mass and the surface gravity of a black hole are identified with entropy, total energy and temperature respectively \cite{Bekenstein_Black_holes_and_entropy,Hawking:1994_Particle_Creation} and their differential changes satisfy a relation which is similar to the first law of thermodynamics. In order to confirm that the black holes are indeed thermal objects, one needs to understand the microscopic origin of the thermal structure. This is still a challenging and \emph{open} problem in black hole physics. String theory has successfully provided a partial answer to this question in the context of supersymmetric extremal asymptotically flat black holes \cite{Strominger:1996sh}. A comprehensive understanding of black hole microscopy for a generic class of black holes, including the  $AdS$ ones, is missing. In this paper, we try to shed some light towards the understanding of $AdS$ black hole microstructure.

A surprising similarity between electrically charged $AdS$ black holes and van der Waals fluid was first observed in \cite{Chamblin:1999tk_charge_BH}. The phase structure of charged black hole in the fixed charge ensemble is similar to that of a van der Waals fluid (non-ideal fluid) if one identifies the inverse temperature, electric charge and horizon radius of the black hole with pressure, temperature and volume of the liquid-gas system respectively \cite{Chamblin:1999tk_charge_BH, Chamblin:1999tk_charge_BH_2}. Such resemblance was further studied by \cite{Kubiznak+Mann:2012wp_PVCriticality,Kubiznak:2014BHchemistry,Kubiznak:2016BH-chem-lambda} with a different identification between the parameters. Following the earlier works \cite{Kastor:2009wy,Sekiwa:2006qj_PLambda1,Caldarelli:1999xj_PLambda2,Wang:2006eb_PLambda3}, Kubiznak and Mann considered the negative cosmological constant to be thermodynamic pressure of the AdS black hole and volume covered by the event horizon to be thermodynamic volume conjugate to pressure\cite{Spallucci:2013adsBH-arealaw, Altamirano:2013_adsBH-triplepoint,Cai:2013P-V_for_GB-BH,Dehyadegari:2016nkd}. This gives an one to one mapping between charges $AdS$ black holes and van der Waals fluid. Although such identification does not explain anything about the microstructure but this was an important observation towards the microscopy of $AdS$ black hole.

Wei and Liu proposed that the thermal $AdS$ black holes have microstructure called \emph{black hole molecules} \cite{Wei+Wen:2015iwa}. Their claim was based on the similarity between the black hole phase structure and that of a van der Waals fluid. Identifying the horizon radius with the specific volume, Wei and Liu introduced the idea of \emph{number density} of black hole molecules (the inverse of specific volume) to measure the microscopic degrees of freedom. They observed that the number density suffers a sudden change accompanied by a latent heat when the black hole undergoes a phase transition. Calculating the Ruppeiner scalar curvature, they also showed that there is a weak attractive interaction between two black hole molecules. However, the origin of such microstructure is not clear. 

G. Ruppeiner developed a geometric interpretation of a thermal system by constructing a thermodynamic line element (or metric) in a space (thermodynamic manifold) spanned by the thermodynamic variables \cite{Ruppeiner:1979,Ruppeiner:1983zz, Ruppeiner:1995zz}. Following the standard techniques of Riemannian geometry G. Ruppeiner constructed a scalar curvature of the thermodynamic manifold. It turns out that the sign of the scalar curvature signifies the nature of interactions between the molecules of the system. In particular the positive (negative) curvature implies repulsion (attraction) between the constituent molecules\footnote{See appendix \ref{app:ruppeiner} for a detailed discussion.}. The idea of Ruppeiner's construction was first implemented to \emph{BTZ} black hole in \cite{Cai:1998ep_BTZ}. Application of this idea on variety of black holes in \emph{AdS} space can be obtained in \cite{Aman:2003ug,Aman:2005xk,Shen:2005nu,Mirza:2007ev,Aman:2007pp,Chaturvedi:2014vpa,Xu:2020sads-rg}. Wei, Liu and Mann constructed the Ruppeiner scalar for charged $AdS$ black holes in temperature and volume plane and showed that the characteristic curves in $(T, V)$ plane are similar to those of van der Waals fluid\footnote{See also \cite{Wei+R.Mann:2019yvs,Wei:2020poh,Ghosh:2019pwy,Dehyadegari:2020ebz}.} except some interesting corners. There are regions in the $(T, V)$ plane where the sign of Ruppeiner scalar is negative (positive) denoting attractive (repulsive) interaction between black hole molecules. In \cite{Miao:2017fqg_BH_potential} Miao and Xu conjectured that the empirical Lennard-Jones potential is a feasible candidate to provide a qualitative explanation of the interaction between black hole micromolecules. They also showed that the interaction forces due to the Lennard-Jones potential match with that of the Ruppeiner scalar curvature.

In this paper we take a bottom-up approach to find the interaction potential between the black hole molecules. The equation of states of $AdS$ black holes is similar to the van der Waals fluid, which describes the state of interacting particles under mean field approximation. At the same time, the critical exponents for the AdS black hole evaluated on the equation of state also take the ”mean-field” values. A natural question arises at this point whether there is an effective mean field interaction between the black hole molecules of $AdS$ black holes. To answer this question we start with the equation of state of $AdS$ black hole in fixed charge ensemble and use the method of classical cluster expansion \cite{stat_mech_book_Huang,stat_mech_book_Mayer} to find the potential. It turns out that the Lennard-Jones (LJ) potential is a feasible candidate to describe such interaction. Our analysis supports the conjecture proposed in \cite{Miao:2017fqg_BH_potential}. The LJ potential describes a two-body interaction. It has two parameters : strength of the potential $\epsilon$ and the critical distance $r_0$ such that two interacting particles repel each other for $r<r_0$ (i.e. when they come close to each other) and attract for $r>r_0$. We compute the value of $\epsilon$ and $r_0$ for dyonic $AdS$ black holes and compare the result with that obtained from the Ruppeiner scalar curvature. In particular, we first find the interaction potential for a pure $AdS$ black hole. We observe that for pure $AdS$ black hole the depth of the potential is proportional to the temperature and the critical volume $v_0 \sim r_0^3$ is inversely proportional to the temperature. This result is in agreement with \cite{Xu:2020sads-rg,Wei+R.Mann:2019yvs}. We then turn on the electric and magnetic charges (dyonic black hole). Calculation of cluster integrals in the presence of generic electric and magnetic charges is complicated. In order to understand the effect of charges on the interaction between the molecules, we consider a small charge approximation. Our analysis shows that the strength of the potential and the critical volume reduces in the presence of electric (and magnetic) charges.

The paper is organized as follows. 
\begin{itemize}
    \item 
    In Sec. \eqref{Sec:cluster_review}, we review the cluster expansion technique and compute the relation between virial coefficient and irreducible form of the cluster integrals.
    \item
    In Sec. \eqref{Sec:black_hole_review}, we discuss the thermodynamics and equations of state of the Schwarzschild black holes and the dyonic black holes in asymptotic $ AdS $ space-time.
    \item
    In the Sec. \eqref{Sec:correspondence}, we construct the mean-field potential by equating the the cluster integrals with the virial coefficients. It turns out that the Lennard-Jones potential is a feasible candidate to describe the interaction between the black hole molecules.
    \item
    Finally we summarize and discuss the results in \eqref{Sec:conclusion}.
    \item
    In appendix \ref{app:ruppeiner} we present the construction of Ruppenier curvature scalar for dyonic $AdS$ black hole and discuss the relation between the signature of Ruppenier scalar and the nature of interaction between black hole molecules.
    \item
    In appendix \ref{app:clustercal} we compute the first few cluster integrals required in our computation.

\end{itemize}


\section{Cluster Expansion for non-ideal gas}\label{Sec:cluster_review}

Cluster	expansion is a useful technique to compute the virial coefficient for an interacting gas of particles. The method was introduced by Mayer and Ursell. We use this technique to find the mean-field potential for black hole molecules. In this section, we briefly review the method. For details follow \cite{stat_mech_book_Kardar_M, stat_mech_book_Huang, stat_mech_book_Mayer}.

Consider a system of $N$ interacting particles of mass $m$ in volume $V$ and at temperature $T$. The Hamiltonian of the system is given by 
\begin{equation}\label{vdw hamiltonian}
	\begin{aligned}
		\mathcal{H} = \sum_{i = 1}^{N} \frac{\mathbf{p}^2_i}{2m} + \mathcal{V}(\mathbf{q}_1, \mathbf{q}_2, \dots , \mathbf{q}_N) \, .
	\end{aligned}
\end{equation}
Here $\mathbf{p}_i$ is the momentum of $i^{th}$ particle and the total potential $\mathcal{V}$ is given by
\begin{equation}
	\mathcal{V}(\mathbf{q}_1, \mathbf{q}_2, \dots , \mathbf{q}_N)  = \sum_{i < j}^{} v_{ij}(\mathbf{q}_i - \mathbf{q}_j)
\end{equation}
where $v_{ij}$ is the potential between $i^{th}$ and $j^{th}$ particles which depends on their positions $\mathbf{q}_i$ and $\mathbf{q}_j$. Furthermore, the grand canonical partition function is given by
\begin{widetext}
	\begin{eqnarray}
		\begin{aligned}
			{\cal L}(z,V,T) & = \sum_{N=0}^\infty z^{N} Q_N(V,T) \\
			\text{where}\quad	Q_N(V,T) & = \frac{1}{N! \, \hbar^{3N}} \int d^{3N} \mathbf{p} \; d^{3N} \mathbf{q} \; \exp \left\{  -\beta \sum_{i = 1}^{N} \frac{\mathbf{p}^2_i}{2m} -\beta\sum_{i < j}^{} v_{ij}(\mathbf{q}_i - \mathbf{q}_j) \right\}
		\end{aligned}
	\end{eqnarray}
\end{widetext}
where $\beta = \frac{1}{k_B T}$ is the inverse temperature and $z$ is the fugacity. The integral over the momenta can be done and the final result is given by,
\begin{equation}
	\label{eq:pfposition}
	\begin{split}
		{\cal L}(z,V,T) & = \sum_{N=0}^\infty \left( \frac{z}{\lambda^3} \right)^N \frac{Z_N(V,T)}{N!} \\
		\text{where} \quad	Z_N(V,T) & = \int d^{3N}q \exp\left( -\beta \sum_{i<j} v_{ij}\right).
	\end{split}
\end{equation}
Here $\lambda = \sqrt{2\pi \hbar^2 / m k T}$ is the thermal wavelength. The integral in \eqref{eq:pfposition} is known as \emph{configuration integral} (CI). 

CI can be solved by using the technique of classical cluster expansion. Defining a Mayer's function $f_{ij}=e^{-\beta v_{ij}}-1$ the CI can be written as
\begin{equation}
	Z_N(V,T) = \int d^{3N}q \prod_{i<j} (1+f_{ij}).
\end{equation}
In order to solve the above integral one defines an $ \ell $-linked cluster as a connected (by single lines) diagram of $\ell$ nodes. An $\ell$-cluster can not be reduced into smaller clusters without cutting a single line. For example $f_{ij}f_{jk}$ denotes a single cluster of 3 nodes (3-cluster) carrying numbers $i,j$ and $k$ where the node $j$ is connected to both $i$ and $k$ but the later two are not connected. One can also define an $\ell$-linked cluster integral $b_\ell$
\begin{equation}
	\label{linked cluster integral def}
	{b}_\ell = \frac{1}{\ell ! \lambda^{3\ell- 3} V} \int d \mathbf{q}_1 \dots d \mathbf{q}_\ell \Tilde{\sum_{\ell\geq i>j \geq 1}} \prod f_{i j} 
\end{equation}
where, $\Tilde{\sum}$ denotes the sum over all the products of $f_{ij}$ consistent with a $\ell$-cluster as defined above. At thermal equilibrium the thermodynamic quantities pressure and density of the system can be written in terms of cluster integrals as,
\begin{eqnarray} \label{eq:Pandv} 
	\frac{P}{k_B T} =
	\frac{1}{\lambda^3}\sum_{\ell=1}^{\infty} \bbar_\ell z^\ell, \quad \frac{N}{V}
	= \frac{1}{v} = \frac{1}{\lambda^3}\sum_{\ell=1}^{\infty} \ell \bbar_\ell
	z^\ell,
\end{eqnarray}
where
\begin{eqnarray}\label{bbarl}
	\bbar_\ell(T) = \lim_{V \rightarrow \infty }b_\ell(V,T) .
\end{eqnarray}
\\
From eq.(\ref{eq:Pandv}) one can find that the virial expansion of
equation of state is given by,
\begin{eqnarray} \label{eq:virialexp}
	\frac{P v}{k_B T} = \sum_{\ell=1}^{\infty} a_\ell(T) 
	\left( \frac{\lambda^3}{v}\right)^{\ell-1} ,
\end{eqnarray}
where the virial coefficients $a_\ell$'s are determined in terms of cluster integrals from the following identity,
\begin{widetext}
	\begin{eqnarray}
		\left (\bbar_1 z + 2 \bbar_2 z^2 + + 3 \bbar_3 z^3 + \cdots \right) &&
		\bigg[ a_1 + a_2 \left( \sum_{n=1}^{\infty} n \bbar_n z^n \right) + a_3 \left( \sum_{n=1}^{\infty} n \bbar_n z^n \right)^2 
		+ \cdots \bigg] 
		\nonumber\\
		&& = \bbar_1 z + \bbar_2 z^2 + \bbar_3 z^3 
		+ \cdots . \nonumber
	\end{eqnarray}
	First few of them are given by,
	\begin{eqnarray} \label{eos-gas}
		a_1 &=& \bbar_1 = 1, \quad a_2 = -\bbar_2, \quad 
		a_3 = 4\bbar_2^2 - 2 \bbar_3, \quad
		a_4 = -20 \bbar_2^3 + 18 \bbar_2\bbar_3 -
		3 \bbar_4, \quad \cdots .
	\end{eqnarray}
\end{widetext}

\subsection{Irreducible Integral form of Cluster Integral}

The cluster integrals (\ref{linked cluster integral def}) are reducible. This means that the integrals can be written as product of smaller integrals. Therefore one can further define irreducible cluster integrals $\beta_k$
\begin{equation}\label{beta integral def}
	\beta_{k}=\frac{1}{{k} !\,  V} \int \cdots \int d \mathbf{q}_{1} \cdots d \mathbf{q}_{{k}+1} \; \bar{\sum_{k+1\geq i>j\geq 1}}\prod f_{i j} \, ,
\end{equation}
where $\bar \sum$ denotes the sum over all $k+1$-clusters which are more than singly connected. One can not reduce such cluster into smaller one by cutting a single line. Using the definitions of \eqref{linked cluster integral def} and \eqref{beta integral def} the cluster integrals $b_\ell$ can be written in terms of irreducible cluster integrals $\beta_k$. The first few relations are given by\footnote{There exists a generic relation between $b_\ell$ and $\beta_k$. See \cite{stat_mech_book_Mayer}.}
\begin{equation}\label{irred integral relation with cluster}
	\begin{aligned}
		& b_{1}=1, \quad b_{2}=\frac{1}{2} \beta_{1}, \quad b_{3}=\frac{1}{2} \beta_{1}^{2}+\frac{1}{3} \beta_{2}\;,\\
		& b_{4}=\frac{2}{3} \beta_{1}^{3}+\beta_{1} \beta_{2}+\frac{1}{4} \beta_{3} \;, \cdots .
	\end{aligned}
\end{equation}
Finally the equation of state for non-ideal fluid can be written in terms of irreducible integrals
\begin{equation}\label{eos in  term of irred int}
	\frac{P v}{k_B T} = 1 - \sum_{k \geq 1} \frac{k}{k+1} \beta_k \left( \frac{1}{v} \right)^k .
\end{equation}
Comparing (\ref{eq:virialexp}) and (\ref{eos in  term of irred int}) one can find relations between virial coefficients and irreducible cluster integrals
\begin{equation} \label{eq:a-beta-rel}
	a_2 = -\frac12 \beta_1, \quad a_3 = - \frac23 \beta_2, \quad a_4 = - \frac{3}{4}\beta_3 \cdots . 
\end{equation}


\section{Thermodynamics of \texorpdfstring{$AdS$}{ads} dyonic Black Holes and equation of state}\label{Sec:black_hole_review}

To derive the equation of state of a dyonic $AdS$ black hole in $3+1$ dimensions in different ensembles we start with the Reissner-Nordstr\"om action
\begin{equation}\label{action EHM}
	S_{EM} = \frac{1}{16 \pi G}\int d^{4}x \sqrt{-g} \, \left(  \mathcal{R} - \mathcal{F}^2 + \frac{6}{L^2} \right)
\end{equation}
where $\mathcal{R}$ is the Ricci Scalar, ${\cal F}$ is the $U(1)$ field strength and $L$ is the radius of $AdS$ space that is related to the cosmological constant $\Lambda = - \frac{3}{L^2}$. The equation of motion obtained from this action admits a spherically symmetric solution
\begin{align}\label{dyonic BH metric}
	ds^2 = -f(r)dt^2 + \frac{dr^2}{f(r)} + r^{2}\left(d \theta^{2} + \sin^{2} \theta d \varphi^{2}\right),
\end{align}
\begin{equation}
	\begin{aligned}
		\text{where} \quad  f(r) & = 1 + \frac{r^2}{L^2}-\frac{2 M}{r}+\frac{q^2_e + q^2_m}{r^{2}},\\
		A_\mu dx^{\mu} & = \left( - \frac{q_e}{r} + \frac{q_e}{r_+} \right) dt +  ( q_m \cos{\theta} ) d\varphi
	\end{aligned}
\end{equation}
where the integration constants $M$, $q_e$ and $q_m$ carry physical meaning of mass, electric and magnetic charges respectively. $A_\mu$ is the $U(1)$ gauge field and	$r_+$ be the position of the outer horizon : $f(r_+)=0$. The asymptotic value of the time component of the gauge field is considered to be electric potential of the system
\begin{equation}
	\Phi_e \sim \frac{q_e}{r_+}.
\end{equation}

In order to study the thermodynamics and phase structure of the system one first choose an ensemble. One can either choose fixed electric potential and magnetic charge ensemble \cite{Dutta:2013dca,Cheng:1993wp} or fixed electric and magnetic charge ensemble. The Hawking temperature in fixed charge ensemble is given by
\begin{equation}\label{E:thetemperature}
	T=\frac{1}{\beta}=\frac{1}{4\pi r_+}\left[1+
	\frac{3r_+^2}{L^2}-\frac{q_e^2+q_m^2}{r_+^2} \right].
\end{equation}
In our analysis we consider the cosmological constant $\Lambda=-6/L^2$ as thermodynamic pressure of the system which is different than the standard treatment. The idea of considering cosmological constant as thermodynamic pressure was first introduced in\footnote{It was suggested in \cite{Sekiwa:2006qj_PLambda1,Caldarelli:1999xj_PLambda2,Wang:2006eb_PLambda3,Kastor:2009wy} that $\Lambda$ can be treated as a thermodynamic variable of the system.} \cite{Kubiznak+Mann:2012wp_PVCriticality}, i.e. 
\begin{equation}
	P=-\frac{\Lambda}{8\pi} = \frac{3}{8\pi}\frac{1}{L^2} 
\end{equation}
since in presence of a cosmological constant the first law of black hole thermodynamics becomes inconsistent with the Smarr relation unless the variation of $\Lambda$ is included in the first law. Once we consider variation of $\Lambda$ in the first law, the black hole mass $M$ is identified with enthalpy rather than internal energy of the system \cite{Kastor:2009wy,Kubiznak+Mann:2012wp_PVCriticality,Kubiznak:2016BH-chem-lambda}. 

Following \cite{Dutta:2013dca,Balasubramanian:1999re} one can compute the free energy of $AdS$ dyonic black holes in fixed electric charge ensemble
\begin{equation}\label{Eq:BHfreeenergy}
	W=\frac{I}{\beta}=\frac{1}{4 G_4}\left[r_+ + \frac{3(q_e^2+q_m^2)}{r_+} - \frac{8\pi Pr_+^3}{3}  \right] . 
\end{equation}
The other thermodynamic variables are given by,
\begin{equation}\label{thermoquantity}
\begin{split}
	\Phi_e &=\frac{\partial W}{\partial q_e}=\frac{q_e}{G_4 r_+}, \quad
	\Phi_m =\frac{\partial W}{\partial
		q_m}=\frac{q_m}{G_4\ r_+} , \\
	S &=-\frac{\partial W}{\partial T}=\beta^2\frac{\partial W}{\partial
		\beta}=\frac{1}{4G_4}4\pi
	r_+^2=\frac{A_H}{4G_4}, \\ 
	V  & = \frac{\partial W}{\partial
		P}=\frac{4\pi}{3}r_+^3 .
\end{split}
\end{equation}

Using equation (\ref{E:thetemperature}) one can write the pressure as a function of $T$, $r_+$,
$q_e$ and $q_m$
\begin{equation}\label{E:EOS1}
	\frac{P v}{T} = 1-\frac{1}{2\pi T } \cdot \frac{1}{v} + \frac{2(q_e^2+q_m^2)}{\pi T} \cdot \frac{1}{v^3}.
\end{equation}
where, $v=2r_+$ can be identified with the specific volume of the system. The above relation is the equation of state of a dyonic black hole in fixed charge ensemble. One can also work on fixed electric potential ensemble to start with \cite{Dutta:2013dca}. The equation of state in that ensemble takes the form
\begin{equation}\label{E:EOS2}
	\frac{P v}{T} = 1-\frac{1-\Phi_e^2}{2\pi T }\cdot \frac{1}{v} + \frac{2q_m^2}{\pi T} \cdot \frac{1}{v^3}.
\end{equation} 
These equations describe different phases of a dyonic black hole in different ensembles. Both the equations have surprising similarity with the van der Waal equation or virial expansion in general. However there is a potential difference between these equations and virial expansion. Unlike virial expansion both the equations (\ref{E:EOS1} and \ref{E:EOS2}) truncate after third order. Our goal is to study the interaction between black hole molecules which give rise to these equations of state.

\section{Mean field potential for black hole molecules }\label{Sec:correspondence}

Van der Waals equation qualitatively explains the equation of the state of non-ideal fluid. The equation of state and the critical exponents for van der Waals fluid can be derived under a mean field approximation. In mean field theory, we assume that each particle moves independently in an average potential field offered by the other particles. The famous Lennard-Jones potential is a good choice of mean field potential to explain the behaviour of non-ideal fluid qualitatively. The LJ potential has two terms: the repulsive term (proportional to $1/r^{12}$) describes the Pauli repulsion at short distances between the molecules and the attractive term (proportional to $1/r^{6}$) describes attraction at long distances.

The similarity between the equations of state of $AdS$ black holes and van der Waals fluid and also the equality of the critical exponents of the two systems motivate one to postulate a microstructure for $AdS$ black holes \cite{Kubiznak+Mann:2012wp_PVCriticality,Kubiznak:2014BHchemistry,Wei+Wen:2015iwa}. Further, computation of Ruppeiner thermodynamic scalar for charged $AdS$ black holes and van der Waals fluid also shows a surprising similarity between characteristic curves of both the systems in $(T, V)$ plane \cite{Wei:2019uqg, Wei+R.Mann:2019yvs}. The characteristic curves contain information about the attractive and repulsive nature of the interaction between the molecules. Thus similarity of characteristic curves indicates that there could be a similar microstructure for $AdS$ black holes. It has also been conjectured that the LJ potential qualitatively explains the phase structure of $AdS$ black hole \cite{Miao:2017fqg_BH_potential}. In order to understand the underlying microstructure of $AdS$ black holes in more detail, we try to derive the mean field potential between the black hole molecules such that the corresponding equation of state is given by (\ref{E:EOS1}) (or \ref{E:EOS2}). Using the relations between virial coefficients and the irreducible cluster integrals (\ref{eq:a-beta-rel}) we find that the LJ potential is a feasible choice to explain the interaction. The LJ potential has a \emph{critical} length denoted by $r_0$. If the distance between the molecules is greater (less) than that, then the interaction between them is attractive (repulsive). We find that the strength of LJ potential can be adjusted such that the value of the critical length $r_0$ (for a given temperature and charges) matches with the result derived from the zeros of Ruppeiner curvature.

\begin{figure}[H]
	\centering
	\includegraphics[width=1.0\linewidth]{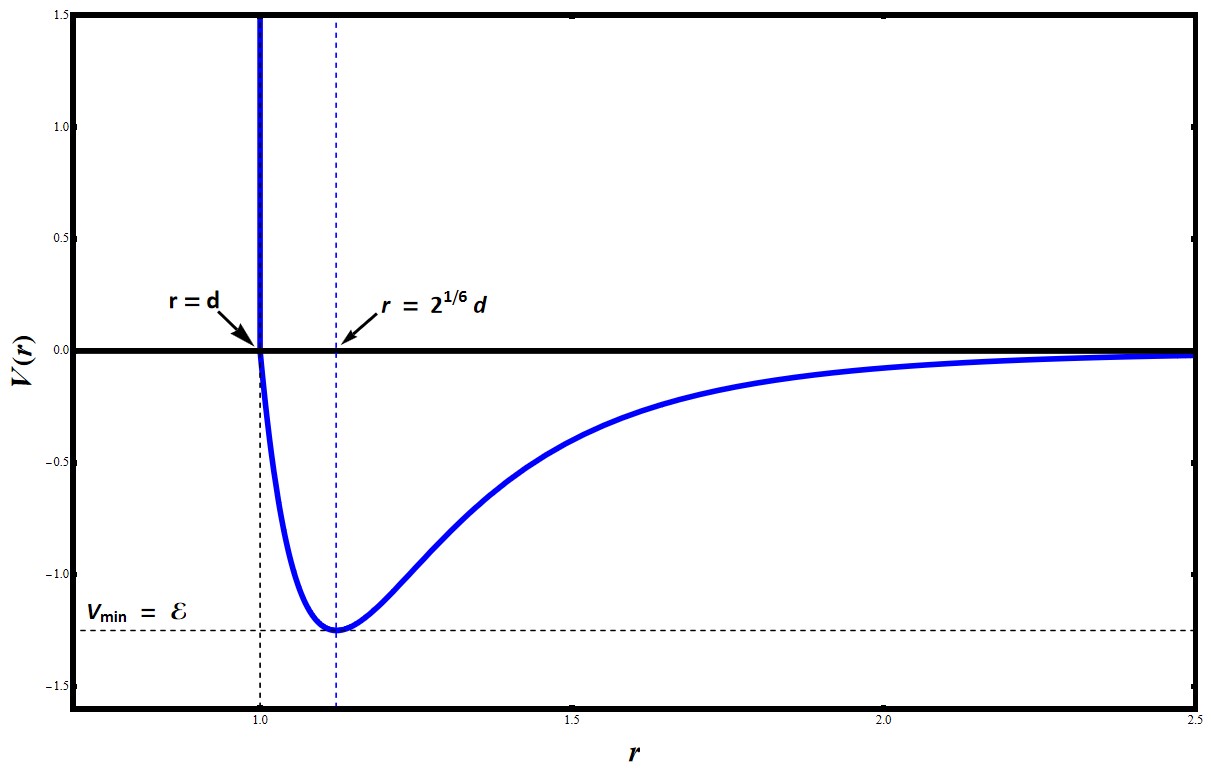}
	\caption[Plot of mean-field potential]{Modified Lennard-Jones potential.}
	\label{fig:modifiedpotentialansatz}
\end{figure}

We find the following form of the potential suitably describes the interaction between $AdS$ black hole molecules\footnote{We modified the LJ potential for small $r$. The Lennard-Jones potential is a very rapidly increasing function of $r$ when $r$ is close to zero (it goes like $\sim 1/r^{12}$). This implies a very strong repulsion between the molecules and hence two molecules coming in the vicinity of $r<d$ is extremely less probable. Therefore the modified potential is approximately equal to the Lennard-Jones potential. One can also work with the actual LJ potential. In that case the dependence of parameter $\epsilon$ on temperature (\ref{eq:epsi-T-dependnce}) would remain same up to a very small change in the numerical factors.}
\begin{align}\label{potential ansatz}
	\mathcal{V}(r)= 
	\begin{cases}
		\infty & \text{for} \  r<d \\
		4 \varepsilon \left[\left(\frac{d}{r}\right)^{12} - \left(\frac{d}{r}\right)^{6} \right]  & \text{for} \ r\geq d.
	\end{cases}
\end{align}
Here $r$ is the distance between two interacting molecules, $\varepsilon$ is the depth of the potential and $d$ is the distance at which the particle-particle interaction potential is zero. $d$ is proportional to the size of the molecules. In order to avoid the singular behaviour at $r=0$ we consider hard sphere approximation i.e. the potential is infinity for $r<d$. This implies that there exists a restricted volume $\omega=\frac{4\pi}{3}d^3$ where no molecules can enter as the repulsion is infinity. For $r>d$ the potential has a minimum at 
\begin{equation}
	r_0=2^{1/6} d.   
\end{equation}
From $r=d$ to $r=r_0$ the potential is repulsive and from $r=r_0$ to $\infty$ it is attractive. The minimum value of the potential is $\varepsilon$.
See figure \ref{fig:modifiedpotentialansatz}.

The Mayer function under mean field approximation takes the following form,
\begin{align}\label{eq:mayerF}
	f_{ij}(r)= 
	\begin{cases}
		-1 ,& \text{for} \  r<d \\
		\exp\left[ -\beta {\cal V}(r) \right]-1               & \text{for} \ r\geq d, \quad \text{for all} \ i,j.
	\end{cases}
\end{align}
We define a volume $v_0$ associated with the critical length $r_0$
\begin{equation}
	v_0=\frac{4\pi}{3}r_0^3.
\end{equation}
Our goal is to compute this volume as a function of temperature and charge and compare the result obtained from the thermodynamic curvature \cite{Wei+R.Mann:2019yvs}.

Computation of (irreducible) cluster integrals for this potential is in general complicated. To get an analytic handle on the problem we make an approximation to simplify the Mayer function. We assume that $\beta {\cal V}(r)$ is small enough such that the Mayer function can be approximated as
\begin{align}\label{eq:mayerF2}
	f_{ij}(r)= 
	\begin{cases}
		-1 ,& \text{for} \  r<d \\
		- 4 \beta \varepsilon \left[\left(\frac{d}{r}\right)^{12} - \left(\frac{d}{r}\right)^{6} \right]  & \text{for} \ r\geq d, \quad \text{for all} \ i,j.
	\end{cases}
\end{align}
In what follows we see that in the high temperature limit this is a good approximation.

Using (\ref{linked cluster integral def}) and (\ref{beta integral def}) we can write
\begin{eqnarray}
\begin{aligned}
	\beta_1	= 2 \bbar_2  & = \frac{1}{V} \iint d^3 \mathbf{q}_1 d^3 \mathbf{q}_2 f_{ij}( \mathbf{q}_1, \mathbf{q}_2 ) \\
	& = - \frac{\beta}{V} \iint d^3 \mathbf{q}_1 d^3 \mathbf{q}_2  \mathcal{V}(r) , \ \text{where} \ r = |\mathbf{q_1}-\mathbf{q_2}|.
\end{aligned}
\end{eqnarray}
Going to the center of mass frame it is easy to do this integration and the final result is given by,
\begin{equation}\label{eq:beta1int}
	\beta_1 = \left(-1 + \frac{8 \varepsilon}{3T}\right) \frac{v_0}{\sqrt2}.
\end{equation}

We first consider a neutral $AdS$ black hole and find the strength of the LJ potential $\varepsilon$ and the critical length $r_0$. The second virial coefficient $a_2$ for neutral $AdS$ black hole is given by (\ref{E:EOS1})
\begin{equation}
	a_2 = - \frac{1}{2\pi T}.
\end{equation}
All other higher virial coefficients are zero. Comparing the second virial coefficient with the irreducible cluster integral $\beta_1$ one can choose 
\begin{equation}\label{eq:epsi-T-dependnce}
\varepsilon = \frac{3}{8} \left(1+2 \sqrt{2}\right) T \sim 1.44 \ T
\end{equation}
such that $v_0$ is given by
\begin{equation}\label{v0Trel}
	v_0 = \frac{1}{2\pi T}.
\end{equation}
In the high temperature limit this calculation is quite reliable since higher order irreducible cluster integrals are suppressed by higher powers of $1/T$ 
\begin{equation}
	\beta_k \sim \frac{1}{T^k} \quad \text{for $k>1$}
\end{equation}
and hence the higher virial coefficients.

Thus we see that the LJ potential is a feasible candidate to describe a mean field interaction between the black hole molecules in the high temperature limit. From the potential, we see that the interaction between two molecules is attractive if the specific volume $v$ is greater than the critical volume $v_0=1/2\pi T$. The result is in agreement with the result obtained from the computation of the Ruppeiner thermodynamic scalar (see appendix \ref{app:ruppeiner}). From the derivation, one can see that $v_0 \sim 1/T$ and $\epsilon \sim T$ up to some numerical values. The numerical value for $\epsilon$ can be adjusted such that (\ref{v0Trel}) holds. Therefore from our calculation we see that the LJ potential (\ref{potential ansatz}) (up to some numerical factors) suitably describes the interaction between black hole molecules for $AdS$ black holes. In the high temperature limit, the argument of the exponential in (\ref{eq:mayerF}) goes as
\begin{equation}
	\beta {\cal V}(r) \sim \frac{1}{T^2} 
\end{equation}
and hence our approximation $f_{ij}(r) \sim -\beta {\cal V}(r)$ is good enough.

\subsection{Dyonic Black Hole}

We next consider dyonic $AdS$ black hole in constant charge ensemble\footnote{The analysis for fixed potential ensemble can be done in similar way.}. The equation of state is given by (\ref{E:EOS1}).
The virial coefficients are given by
\begin{equation}
\label{eq:a1a2a3a4}
   	a_1  = 1, \ a_2 = -\frac{1}{2\pi T}, \
	a_3  = 0, \ a_4 = \frac{2(q_e^2 +q_m^2)}{\pi T}
\end{equation}
and all higher virial coefficients are zero. Calculation of irreducible cluster integrals in presence of arbitrary electric and magnetic charges is tedious. We make a further (apart from the high temperature approximation) assumption here. We take the charges $q_e$ and $q_m$ to be small and keep terms up to order $q_e^2$ and $q_m^2$. We find the perturbative correction to the mean field potential in presence of charges. This helps us to understand how the electric and magnetic charges change the interaction between the black hole molecules.

We correct the ansatz for the Mayer function up to order $q^2$ where $q^2=q_e^2+q_m^2$
\begin{widetext}
\begin{align}\label{eq:mayerFQ}
	f_{ij}(r)= 
	\begin{cases}
		-(1-q^2 \delta) ,& \text{for} \  r<d \\
		- 4 \beta \varepsilon \left[(1+q^2 \gamma )\left(\frac{d}{r}\right)^{12} - (1+q^2 \sigma) \left(\frac{d}{r}\right)^{6} \right]  & \text{for} \ r\geq d, \quad \text{for all} \ i,j
	\end{cases}
\end{align}
\end{widetext}
where $\delta, \ \gamma$ and $\sigma$ are constants depend on temperature and charges. We compute the cluster integrals for the above potential. See appendix \ref{app:clustercal} for details. Using the relations (\ref{eq:a-beta-rel}) we calculate the unknown coefficients $\delta, \gamma$ and $\sigma$. We see that there coefficients can be chosen as follows
\begin{equation}\label{eq:dgs}
	\delta = \frac{2.65}{v_0^2},\quad \sigma = \frac{19.58}{v_0^2} \quad \text{and} \quad \gamma = \frac{36.39}{v_0^2}
\end{equation}
such that the critical volume $v_0$ is given by
\begin{equation}\label{eq:v0_dyonic}
	v_0 = \frac{1}{2\pi T} - 16\pi T q^2
\end{equation}
which is in agreement of thermodynamic curvature calculation (\ref{eq:v0charge}). With these values one can calculate the minimum of the potential it is given by
\begin{equation}
	{\cal V}_{min} = -1.44 \ T \left( 1- \frac{2.78 q^2}{v_0^2}\right)
\end{equation}
up to order of $q^2$.

\section{Discussion}\label{Sec:conclusion}

In this article, we probe the effect of electric and magnetic charges on the interaction between black hole molecules by calculating the effective mean-field interaction potential between them using the techniques of classical cluster expansion for a class of $3 + 1$ dimensional $AdS$ dyonic black holes. The equation of state of charged \emph{AdS} black hole in the extended phase space is similar to that of a van der Waals fluid. To fully understand the microscopic interactions between the black hole molecules, we take a bottom-up approach to find the mean-field potential between them. It turns out that the Lennard-Jones potential is a feasible choice to explain the interaction between the molecules and it matches with the result obtained by calculating the Ruppeiner thermal curvature. 

We first find the LJ potential for a neutral $AdS$ black hole. The computation of irreducible cluster coefficients, in general, is extremely complicated analytically. We consider the high temperature limit and compute these coefficients to find the mean-field potential. It turns out that if we choose the strength of the potential appropriately (the numerical factor), then the critical volume $v_0$ becomes $1/2\pi T$. This implies that if the two black hole molecules come in a volume $v_0$ they start repelling each other; otherwise, there is an attraction between them. This result is in agreement with \cite{Xu:2020sads-rg}. Our next goal was to understand what is the effect of the charge parameter $q$ at the microscopic level. Extending our approach for a generic charge parameter is again a formidable task to do. Therefore we make a further approximation. However, this allows us to understand the effect of charge on the microscopic interaction. It turns out that the charge $q$ reduce the size of the critical volume as well as the strength of the LJ potential. Correcting the Mayer (\ref{eq:mayerFQ}) function for $r<d$ corresponds to deforming the hard sphere approximation in presence of charges. The interaction potential for $r<d$ goes as $T \log\left(1/q^2\right)$. In the limit $q\rightarrow 0$ the potential becomes $\infty$ for $r<d$.

The Ruppeiner curvature scalar for dyonic black holes is given in appendix (equation \ref{ruppenier scalar TV-dyonic}). From the expression we see that the curvature has another zero at $v_0\sim q^2$. Because of this extra zero the characteristic curves studied in \cite{Wei:2019uqg} is different than the characteristic curves for a van der Waals fluid. However in our analysis we do not see this extra solution for the critical volume $v_0$. This is because we studied the charged system as a perturbation over pure $AdS$ black hole. In the limit $q\rightarrow 0$ the second solution is $v_0 = 0$ and it did not show up in our calculation. An exact $q$ calculation might show the second solution for $v_0$.

From the microscopic point of view, different virial coefficients (\ref{eq:virialexp}) account for multi-particle interactions. The $n^{th}$ virial coefficient can be calculated from the $n$-body interaction of the micro molecules. From the equation of state of charged $AdS$ black hole (\ref{E:EOS1} or \ref{E:EOS2}) we see that the micro-structure for $AdS$ black hole is potentially different from the standard non-ideal gas. The first term in the equation of state represents the ideal gas part. At a very high temperature ($T\rightarrow \infty$) one can approximate the black hole molecules to be free gas. The second term in the equation implies that at finite temperature, the underlying molecules are not like ideal gas; rather, there is a repulsive two body interaction between them. For a neutral $AdS$ black hole, there are no multi-body interactions between the molecules. Whereas for charged-$AdS$ black hole there is only two body and four-body interactions at the microscopic level where the charge parameter $q$ accounts the strength of the four body interaction. From our analysis, we see that the four body interaction is attractive in the limit of a small charge. The higher order virial coefficients appear under consideration of higher derivative terms in the action in higher dimensions \cite{Ghosh:2019pwy,Aman:2005xk,Cvetic:2001bk,Wei:2020poh}. It would be interesting to study the effect of higher derivative terms on the interactions between the molecules. Our analysis can also be extended to rotating black holes also. From the temperature of Kerr-AdS black holes one can write the equation of state and read the virial coefficients. It would be interesting to extend this computations for rotating black holes and compare the result with the same from Ruppeiner scalar computation \cite{Aman:2005xk,Wei:2021lmo}.

Probing the micro-structure via the equation of state does not allow us to count the number of microstates to match with the classical entropy. Consideration of black hole molecules is in some sense ad-hoc. We presume that there exists some microstructure, but we exactly do not know the origin of such states. String theory, on the other hand, has been able to give a partial answer in order to construct the microstructure for macroscopic black holes. However, such black holes solutions are asymptotically flat and have a large amount of supersymmetry in order to compute the indices in the weak coupling limit \cite{Strominger:1996sh}. An attempt to compute the microstructure for non-extremal black holes can be found in  \cite{Sfetsos:1997xs,Horowitz:1996ay,Breckenridge:1996sn}.  Computation of indices of supersymmetric field theories also allows one to understand the underlying microstates for the dual black holes in the context of the $AdS/CFT$ correspondence. It would be interesting to explore if there is any connection between the ad-hoc microstructure for supersymmetric $AdS$ black holes and microstructures computed from supersymmetric indices.

\vspace{.5 cm}

\noindent
{\bf Acknowledgement}\ The current work is the completion of an unpublished paper \cite{Dutta:2016urd_mean_field} by one of the author. We would like to thank Robert Mann for useful discussion. The work of SD is supported by the \emph{MATRICS} grant (no. \emph{MTR/2019/000390}, the Department of Science and Technology, Government of India). We are indebted to people of India for their unconditional support toward the researches in basic science.


\appendix
\begin{appendices} 

\section{Ruppeiner Geometry}\label{app:ruppeiner}

Ruppeiner introduced a geometric interpretation of the thermal system introducing the thermodynamic line element in a space spanned by the macroscopic variables \cite{Ruppeiner:1979, Ruppeiner:1983zz, Ruppeiner:1995zz}. The line element defines the geometry of a manifold of thermodynamic variables. A thermodynamic system in equilibrium is expressed in terms of a set of macroscopic variables $\{x^\mu\}$. The infinitesimal distance between two neighbouring thermal states is given by a thermodynamic line element. Ruppeiner showed that the several thermodynamic properties of the system were significantly encoded in the geometry of the thermodynamic manifold. The information of the phase structure of the system is captured by the Riemann scalar curvature of the thermodynamic manifold. Applying this idea to ideal Bose, Fermi, and classical ideal gas, one finds that the corresponding scalar curvature is positive, negative, and zero, respectively \cite{Janyszek_1990_quantum_gases}. This observation suggests that the sign of scalar curvature can be used to understand the nature of interactions between the constituent particles of a thermodynamic system \cite{Ruppeiner:1995zz,Janyszek_1990_quantum_gases, Janyszek_1990}. This is, therefore, in some sense, a bottom-up approach in contrast to the standard statistical mechanics.

In this appendix, we provide a summary of Ruppeiner's idea. Ruppeiner geometry represents the covariant construction of the fluctuation theory of equilibrium thermodynamics. Fluctuation theory states that the neighbouring fluctuation depends on the nearby thermodynamic parameters. The combination of the axioms of thermodynamic and fluctuation theorem leads to the line element $ \Delta l^2 $ for the distance between two neighbouring fluctuations in thermal parameter space is given by
\begin{equation}\label{thermal line element}
	\Delta l^2 = - \frac{\partial^2 S}{\partial x^\alpha \partial x^\beta} \bigg|_{x_0} \Delta x^\alpha \Delta x^\beta = - g_{\alpha \beta} \Delta x^\alpha \Delta x^\beta \;,
\end{equation}
where $ g_{\alpha \beta} = - \frac{\partial^2 S}{\partial x^\alpha \partial x^\beta} \big|_{x_0} $ and $ x_0 $ be the equilibrium point for thermal parameters and $ g_{\alpha \beta} $ be the metric in thermal manifold in term of thermodynamic entropy function $ S(x, x_0) $. Then the line element \eqref{thermal line element} in thermal manifold is given by 
\begin{align}\label{thermal line element 2}
	\Delta l^2 = \frac{1}{T} \Delta T \Delta S - \frac{1}{T} \Delta P \Delta V + \sum_{i} \frac{1}{T} \Delta \mu_i \Delta N^i \, .
\end{align}
\subsection{Ruppeiner geometry for  Black holes}\label{Sec:black_hole_thermalgeometry}

Here we apply the Ruppeiner geometry for the AdS black hole where thermal parameter space constructed with $ (T, S, P, V) $, where $ T $ be the Hawking temperature, $ S $ be the entropy of black hole, $ P $ be the pressure of AdS black hole associated with the cosmological constant and $ V $ be the thermodynamic volume of the black hole. For $ (T, V) $ fluctuation parameter space, the Helmholtz free energy $ F $ be the governing potential for the thermal system, Helmholtz free energy can be calculated as $ F = W - T S $, where $ W $ be the Gibbs free energy given in Eq. \eqref{Eq:BHfreeenergy}.

The line element in Eq. $ \eqref{thermal line element 2} $ for $ (T, V) $ fluctuation can be written as
\begin{equation}\label{line element in TV variable}
	\begin{aligned}
		\Delta l^2 = - \frac{1}{T} \left( \frac{\partial^2 F}{\partial T^2} \right) \Delta T^2 + \frac{1}{T} \left( \frac{\partial ^2 F}{\partial V^2} \right) \Delta V^2.
	\end{aligned}
\end{equation} 
As the specific heat at constant volume is define as $ C_V = T \left( \frac{\partial S}{\partial T} \right)_{V} = - T \left(\frac{\partial^2 F}{\partial T^2}\right)_V $, so metric in term of specific heat is 
\begin{equation}\label{modified metric in TV}
	\Delta l^2 =  \frac{C_V}{T^2} \Delta T^2 - \frac{(\partial_{V} P)_T}{T} \Delta V^2
\end{equation} 
The specific heat at constant volume $ C_V $ of the AdS black hole vanishes. As $ C_V = T \left( \frac{\partial S }{\partial T} \right)_V $ and for (3+1)-dimensional black hole case $ S \propto  A_{bh} \propto  r{\! _{_h}}^{2} $ and $ V \propto r{\! _{_h}}^3 $, which implies $ \left( \frac{\partial S}{\partial r{\! _{_h}}} \right)_V = 0 $. The vanishing heat capacity $ C_V $ implies that the temperature component of line element in thermal geometry for black hole $ g_{TT} = 0 $ or $ g^{TT} = \infty $. So we take vanishing heat capacity as the limit $ k_B \rightarrow 0^+ $ . So treating $ C_V $ as constant such that $ C_V \rightarrow 0^+ $. The scalar curvature diverge as specific heat tends to zero, so we modify the scalar curvature such that
\begin{equation}\label{modified currvature}
	\mathcal{R}_N = \mathcal{R} \, C_V
\end{equation}
Modified Ricci scalar for Eq. $ \eqref{modified metric in TV} $ geometry in term of Hawking temperature and thermodynamic volume $ V $ for Schwarzschild black hole and dyonic black hole is give by  
\begin{widetext}
\begin{subequations}
	\begin{empheq}[]{align}
		\mathcal{R}^{(TV)}_{SAdS} & = \frac{ 1 - 2 \sqrt[3]{6} \pi ^{2/3} T V^{1/3}}{2 \left(\sqrt[3]{6} \pi ^{2/3} T V^{1/3} - 1 \right)^2} \label{ruppenier scalar TV-SAdS} \;, \\
		\mathcal{R}^{(TV)}_{\text{\tiny{Dyonic}}} & = \frac{\left(8 (q_e ^2 + q_m^2) - \left(\frac{6}{\pi }\right)^{2/3} V^{2/3}\right) \left(8 (q_e ^2 + q_m^2) + 12 T V - \left( \frac{6}{\pi } \right)^{2/3} V^{2/3}\right)}{2 \left(8 (q_e ^2 + q_m^2) + 6 T V - \left( \frac{6}{\pi } \right)^{2/3} V^{2/3}\right)^2} \label{ruppenier scalar TV-dyonic}
	\end{empheq}
\end{subequations}
\end{widetext}

Solve the Ruppenier curvature for their vanishing point by inserting the thermodynamic volume $ V $ in terms of specific volume $ v $. The relation between thermodynamic volume and specific volume is $ V = \pi v^3/6 $. First solve  $ \mathcal{R}^{(TV)}_{SAdS} = 0 $, which gives
\begin{equation}\label{eq:v0sads}
	v_0 = \frac{1}{2 \pi T}
\end{equation} 
and then solve $ \mathcal{R}^{(TV)}_{Dyonic} = 0 $ for small charge, we get
\begin{equation}\label{eq:v0charge}
	v_0 = \frac{1}{2\pi T} - 16\pi T q^2
\end{equation}
where $ q^2 = q_e^2 + q_m^2 $.

\section{Computation of cluster integrals}
\label{app:clustercal}

A general definition of irreducible integral is given in Eq. \eqref{beta integral def}. Here we show the result for the $ 2^{nd} $ and $ 3^{rd} $ irreducible integrals, which we calculated for the the potential ansatz given in Eq. $ \eqref{potential ansatz} $. Explicitly $ \beta_{2} $ integral is written as
\begin{equation}\label{beta2def}
	\beta_{2} = \frac{1}{2! V} \iiint f_{32} f_{31} f_{21} d \mathbf{q}_{1} d \mathbf{q}_{2} d \mathbf{q}_{3}\;,
\end{equation}
where $ f_{ij} = f(|\mathbf{q}_i - \mathbf{q}_j|) = e^{- \beta v_{ij}(r) } - 1 $ is the Mayer's function defined in \eqref{eq:mayerF}. Under mean field approximation, assuming  $ \beta \varepsilon \ll 1 $ for Eq. $ \eqref{potential ansatz} $, we get 
\begin{equation*}
	e^{- \beta v_{ij}(r) } - 1 \approx - \beta v_{ij} (r)
\end{equation*}
To simplify the integral, we first change the integration variables such that $ \mathbf{x} = \mathbf{q}_1 - \mathbf{q}_2, $ $ \mathbf{y} = \mathbf{q}_1 - \mathbf{q}_3, $  $ \mathbf{z} = \frac{ \mathbf{q}_1 + \mathbf{q}_2 + \mathbf{q}_3}{3} $,  which implies $  \mathbf{q_2} - \mathbf{q}_3 = \mathbf{x} - \mathbf{y} $. Then the integral measure becomes 
\begin{align*}
	d \mathbf{q}_{1} d \mathbf{q}_{2} d \mathbf{q}_{3} = d \mathbf{x} \, d \mathbf{y} \, d \mathbf{z}
\end{align*}
We can easily integrate out the independent factors obtained from the change of variables. This reduces the integral in \eqref{beta2def} to a double integral of the form
\begin{align} 
	\beta_{2} = \frac{1}{2} \iint f(\|\vec{x}\|) \; f(\|\vec{y}\|) \; f(\|\vec{x} - \vec{y}\|) \; d \vec{x} \, d \vec{y}
\end{align}

Following Mayer's function from Eq. \eqref{eq:mayerF} along with the mean-field approximation, we can compute the integral as follows:
\begin{equation}\label{beta2 integral value}
	\beta_{2} \approx \frac{\left( \omega \right)^2}{2} \left( - 1 - 2.44 \frac{ \varepsilon^2}{T^2}+ 3.25 \frac{\varepsilon^3}{T^3}\right)
\end{equation}

Now, we can perform a similar simplification for the $ \beta_{3} $ integral,
\begin{equation}
	\begin{aligned}
		\beta_{3}= \frac{1}{6 V} \iiiint \left(3 f_{43} f_{32} f_{21} f_{41}  + 6 f_{43} f_{32} f_{21} f_{41} f_{31} \right. \\ 
		\left. +  f_{43} f_{32} f_{21} f_{41} f_{31} f_{42}\right) d \mathbf{q}_{1} d \mathbf{q}_{2} d \mathbf{q}_{3} d \mathbf{q}_{4}
	\end{aligned}
\end{equation}
using the change of variables: 
\begin{align*}
	\mathbf{x} = \mathbf{q}_1 - \mathbf{q}_2,  \; \mathbf{y} = \mathbf{q}_2 - \mathbf{q}_3 , \; \mathbf{z} = \mathbf{q}_1 - \mathbf{q}_4 \;, \\
	\mathbf{w} = \frac{\mathbf{q}_1 + \mathbf{q}_2 + \mathbf{q}_3 + \mathbf{q}_4}{4}
\end{align*}
which gives us $ d \mathbf{q}_{1} d \mathbf{q}_{2} d \mathbf{q}_{3} d \mathbf{q}_{4} = d \mathbf{x} \, d \mathbf{y} \, d \mathbf{z} \, d \mathbf{w} $.
Again, integrating out the independent factors in integrand, we get
\begin{align*}
	\begin{aligned}
		\beta_{3} = \frac{1}{6} \iiint d \mathbf{x} \, d \mathbf{y} \, d \mathbf{z} \; f{(| \mathbf{x}|)} \; f{(| \mathbf{y}|)} \; f{(| \mathbf{z}|)\; f{(| \mathbf{x} + \mathbf{y} - \mathbf{z}|)}} \\ \left\{ 3 + \; 6 \; f(| \mathbf{x} + \mathbf{y}|) + f(| \mathbf{x}  + \mathbf{y}|)\; f(| \mathbf{x} - \mathbf{z}|) \right\}
	\end{aligned}
\end{align*}
Substituting the value of Mayer's function from Eq. \eqref{eq:mayerF} and assuming the approximation $ x \gg d $, we can write $ |x+y-z| \; \approx \; |x| ,\; |x+y| \; \approx \; |x| , \; |x-z| \; \approx \; |x|, $. Similarly, assuming $ y \gg  z \gg d $, we can approximate $ |x+y-z| \; \approx \; |y| \, , \quad |x+y| \; \approx \; |y| ,\; |x-z| \; \approx \; |z| $.
Using the potential ansatz from Eq. $ \eqref{potential ansatz} $ and working in the mean field approximation, the $ \beta_{3} $ integral is evaluated as:

\begin{equation}\label{beta3 integral value}
	\begin{aligned}
		\beta_{3} \approx \frac{ \omega^3 }{3} \left(-1 + 1.83 \frac{\varepsilon ^2}{T^2} - 10.2 \frac{ \varepsilon^3}{ T^3} - 5.28 \frac{\varepsilon^4}{ T^4} \right. \\ 
		\left. + 14.5\frac{\varepsilon^5}{ T^5} + 1.82\frac{\varepsilon^6}{ T^6}\right).
	\end{aligned}
\end{equation}

\end{appendices}


\bibliographystyle{unsrt}
\bibliography{revision}

\end{document}